\documentstyle[12pt]{article}

\def\pa{\partial}

\begin{document}

\begin{flushright} 
OCHA-PP-76 \\
NDA-FP-24 \\
March 1996
\end{flushright}

\begin{center}
{\Large\bf Stringy and Membranic Theory of Swimming of Micro-organisms}
\footnote{The invited talk given by A. Sugamoto}\\
\vspace{1cm}
M. Kawamura, ${\rm S.\  Nojiri}^{\sharp}$ and A. Sugamoto\\
 \it Ochanomizu University, Tokyo, Japan,\\
 National Defence Academy, Yokosuka, ${\it Japan}^{\sharp}$
\end{center}
\vspace{1cm}
\rm

\begin{abstract}
When the swimming of micro-organisms is viewed from the string and 
membrane theories coupled to the velocity field of the fluid, a number 
of interesting results are derived; 
1) importance of the area (or volume) preserving algebra, 
2) usefulness of the $N$-point Reggeon (membranic) amplitudes, and of the 
gas to liquid transition in case of the red tide issues, 
3) close relation between the red tide issue and the generation of 
Einstein gravity, and 
4) possible understanding of the three different swimming ways of 
micro-organisms from the singularity structure of the shape space.
\end{abstract}

\vspace{1cm}

We are very happy to present our recent works on the swimming of 
micro-organisms at Asia Pacific Conference on Gravitation and Cosmology 
held at Sheraton Walker Hill Hotel in Seoul.  
The talk will be based mainly on Refs.\cite{mi1}-\cite{mi4}. 
Amazing is the fact that there are only three kinds of swimming ways of 
micro-organisms; (1) Ciliated motion for the Paramecium etc. using cilia 
(small hairs covering its surface), (2) flagellated motion for the sperm 
etc. using the flagellum (whip), and (3) the swimming of bacteria with 
the so-called bacterial flagella (whips moving like the wine-opener).
Why does such a simple classification exists?  
That is a very difficult issue, 
but we will challenge it.  Our strategy is to consider the boundary of 
the micro-organisms as the strings or membranes 
$X^{\mu}(t; \xi)$ with the parametrization $\xi$,
whose motion is coupled with the external fluid dynamics.  

\begin{enumerate}
\item The fluid dynamics of the micro-organisms is very simple, 
since the small size $L \ll 1 {\rm mm}$ of the micro-organisms implies 
the Reynolds number $R\ll 1$ and we have the following equations of motion 
for the velocity field $v^{\mu}(x)$ and the pressure $p(x)$:
$\Delta v_\mu = \pa_\mu p/\mu$ or $\Delta (\pa_\mu v_\nu -\pa_\nu v_\mu)=0 $,
where $\mu$ is the coefficient of viscosity.

The deformation of the boundary by the micro-organisms themselves should 
be equal to the fluid velocity there: 
$v^{\mu}(\vec{X}(t; \xi))= \dot{X}^{\mu}(t; \xi)$.
These simple equations can determine the swimming motion of the 
micro-organisms 
\cite{wil1}: Deformation of the micro-organisms by their own ways may cause 
unwanted flows at spacial infinity.  Then the swimming motion is so 
determined as to cancel these flows.  

\item We have analyzed flagellated as well as the ciliated motions 
in the spacial dimensions $D=2$, perturbatively with the small 
deformations $\alpha(t, \theta)$, where $\theta$ is the parametrization 
of the circle (the boundary of the ciliate) and the circle obtained by 
the Joukowski transform from the line (the boundary of the flagellate). 
We have selection rules, i.e., the relation between two required Fourier 
mode-numbers, $n_1$ and $n_2$, in order for the  micro-organisms to swim 
(see Ref.\cite{mi1}).

\item The boundaries of the ciliate and the flagellate for $D=2$ are the 
closed and the open strings, respectively. The selection rules, however, 
differ from the mode number relations in the translation and rotation 
operators (belonging to the Virasoro algebra) of the string theories.  
To understand the discrepancy, it may be helpful to study the area (volume) 
preserving diffeomorphisms, since the incompressibility of the fluid 
implies these. They are the $w_{1+\infty}$ algebra and its generalization 
for $D=2$, the representation theory of which as well as of their quantum 
version are well studied recently.  Quantum version gives the uncertainty 
principle between the complex coordinate $z$ and its canonical conjugate 
$\bar{z}$.  
The Planck constant $\hbar$ in our case can be the thermal fluctuation 
squared of the position, namely,  $\sim (\Delta z)^2$.  In the near future 
a clever application of the representation theory of these algebras will  
classify the swimming ways of micro-organisms.

\item The action of the $N$-body micro-organisms coupled with the 
fluid can be written as follows:
\begin{eqnarray}
 \hat{S}_N & = & \sum_{i=1}^N \int dt \int d^{D-1}\xi_{(i)}
P_{\mu}^{(i)}(t; \xi_{(i)})
\left[ \dot{X}^{\mu}_{(i)}(t; \xi_{(i)}) - v^{\mu}(X_{(i)})
 \right] \nonumber \\
& &-{1 \over 2\pi\alpha'} \int d^D x \left[ {1 \over 4} \omega_{\mu\nu}
\omega^{\mu\nu} -{1 \over \mu} p(x) \pa_{\mu} v^{\mu}
\right], \nonumber
\end{eqnarray}
where $\omega_{\mu \nu}=\partial_{\mu}v_{\nu}-\partial_{\nu}v_{\mu}$ 
is the vorticity, and $P^{(i)}_{\mu}$ is the Lagrange multiplier.  
The system is nothing but the Landau gauge QED with the velocity field $v^{\mu}$ as a gauge field.  On the other hand \lq\lq Landau-Lifshitz hydrodynamics"  
tells us that the temporal growth of the entropy $\dot{S}$ can be given by
\[ \dot{S} = \int d^D x \left[ {\mu \over 2T} (\pa_\nu v_{\lambda} +
\pa_{\lambda}v_{\nu})^2 + {\bf q} \cdot \nabla \left( {1 \over T} \right)
\right],
\]
where ${\bf q}$ is the heat current.
The path integral weight given by the action and the statistical weight 
given by the entropy coincides if the Regge slope $\alpha' $ satisfies                  ${1 \over 4\pi\alpha'}={\mu t^{\ast} \over k_B T}$, 
where the $t^{\ast}$ is the typical time scale or the period of the 
swimming motion \cite{mi3}.

\item Probability $G_{N}$ of having the given positions 
$X_{(1)},\cdots, X_{(N)}$ and the given deformation velocities 
$\dot{X}_{(1)},\cdots, \dot{X}_{(N)}$ of the $N$ micro-organisms, 
can be estimated with the path integral weight just discussed 
above, leading to \cite{mi1}
\begin{eqnarray}
   \lefteqn{ G_N (X_{(1)},\dot{X}_{(1)}; \, \cdots; \, X_{(N)}, 
   \dot{X}_{(N)}) }      \nonumber        \\
   &  & = \int {\cal D} P_{\mu}^{(i)} \, \exp \left\{
          i \sum_{i=1}^{N} \int dt \, \int d^{D-1}\xi_{(i)} \, 
          P_{\mu}^{(i)}                \dot{X}^{\mu}  \right\}       
          \nonumber     \\
   &  & \; \times  \:  \exp \left[ 2 \pi i \alpha' \times \frac{1}{2}
          \sum_{i, j} \int dt_{(i)} \, \int d^{D-1}\xi_{(i)} \, 
           \int dt_{(j)} \, \int d^{D-1}\xi_{(j)}  \right.    
           \nonumber    \\   
    &   &  \; \times \left.  P_{\mu}^{(i)}(t_{(i)}; \xi_{(i)} )               
          G_{\perp}^{\mu \nu} \left( X_{(i)}(\xi_{(i)}) - X_{(j)}(\xi_{(j)})
          \right)  P_{\mu}^{(j)}(t_{(j)}; \xi_{(j)} )  \right].
\nonumber
\end{eqnarray}
The result is the $N$-point Reggeon for $D=2$ 
(field theoretical membrane for $D=3$) amplitude with the 
transverse Green's functions $G_{\perp}$. 
In the terminology of the string or the membrane theory, 
$v^{\mu}$ belongs to the target space whereas the coordinates 
$x^{1}, \cdots, x^{D}$ belong to the parameter space. 
Therefore the existence of the handles may corresponds 
to those of the vortices. If  $\vec{X}_N \rightarrow \infty$,  
$-\dot{\vec{X}}_N $ represents the collective motion of 
$N-1$ micro-organisms and $G_{N}$ is its probability, so that 
the averaged collectively swimming velocity reads
\[ \vec{V}_{N-1} = 
  -\langle \dot{X}_N \rangle = -\sum_{\dot{X}_N} \dot{X}_N 
    G_N (X_1, \dot{X}_1; \cdots ; X_N, \dot{X}_N).
\]

\item In summer, the abnormally generated plankton (micro-organism) 
forms the red tide.  It behaves like a fluid (we gave the illustration 
picture of the red tide at the conference.).  To understand the gas to 
liquid transition according to the increase of the number density of the 
plankton, we have to show the van der Waals type equation of state,
\[ \left( p + {N^2 \over V^2}(a+a'/k_B T + \cdots) \right)
(V-Nb) = N k_B T,
\]
where terms with the coefficients $a, a',\cdots $ represents 
the 2 body attractive forces between atoms (micro-organisms), 
whereas the $b$ represents the occupied volume of the atom (micro-organism).  
To estimate the coefficients $a, a', \cdots $ and $b$ 
we extracts the interaction potential $\phi(x)$ from the action and calculate 
the so-called Meyer integral.  Then we have 
\[ \int d^D x \left( e^{-\phi(x)/k_B T}-1 \right)
= -2b + 2a(k_B T)^{-1} + 2a'(k_B T)^{-2} + \dots.
\]
Vanishing of $a$ occurs by the angle averaging of the multipole type potential 
$ \phi(x) \sim \cos(2 \ell \theta+\alpha)/r^{-2 \ell -2}\  
 (\ell =-2, -3, \cdots )$ between two micro-organisms with the 
relative coordinates $(r, \theta)$ and the same oscillation mode $\ell$. 
Now we find that the gas to liquid transition occurs when the number 
density $n$ of the plankton increases as  $n\geq 10\, 
{\rm plankton}/{\rm m}^3 $ 
for the 0.1 mm plankton in the water at temperature  $T\sim 300 {\rm K} $.  
The result is not so unrealistic!? \cite{mi3}

\item As a similar phenomenon to the 
condensation of micro-organisms (red tide), 
we can consider that the condensation of strings may trigger the phase 
transition of the 2-form gravity to the Einstein gravity.  The 2-form 
gravity has the extra Kalb-Ramond symmetry which is, however,  broken 
by the ansatz of introducing the metric.  Instead, we write down the 
Kalb-Ramond invariant action by introducing the extra string field 
whose condensation leads to the ansatz \cite{2f1}. 
This is a kind of the Meissner 
effect in the string theory.  To understand the gravity from the swimming 
of microorganisms it is worth remembering that 20 years ago Professor Nambu 
called Kalb-Ramond theory the relativistic hydrodynamics.

\item Finally we come to the recent work \cite{mi4} by one of the authors 
(Masako Kawamura). She has calculated for the flagellate the efficiency 
$\eta$ of the micro-organism's swimming $\grave{a}\ la$ Shapere and Wilczek.  
Her result is $\eta = {t^{\ast} \over 16 \pi\mu}{1 \over |\sin \beta'|}$      
   which can be unboundedly large for $\beta' \rightarrow 0$, 
whereas the efficiency of the ciliated swimming is bounded from 
above as        $\eta={\sqrt{5} t^* \over 8 \pi^2 \mu}|\sin \beta|$. 
Therefore the flagellated swimming is much more efficient than the ciliated. 
This might be a reason why the size of the flagellate can be smaller than 
the ciliate. Furthermore, this gives a very interesting indication that in 
the shape space (in which each point represents a different shape of the 
micro-organism), there exists the conical singularity at the point of the 
flagellate.  This can be understood as follows:  One cycle of the swimming 
motion draws a closed path in the shape space and the efficiency is the 
period times the ratio of the area inside the closed path over its perimeter 
squared. 

\item Conclusion at Seoul:
\begin{enumerate}
\item Swimming of micro-organisms can be viewed from string and membrane 
theories, coupled to the velocity field of the fluid.
\item In the future, area (or volume) preserving algebra 
($w_{1+\infty}$ or $W_{1+\infty}$) may help the understanding 
of micro-organisms' swimming.
\item Collective motion of micro-organisms can be studied by the $N$-point 
Reggeon (membranic) amplitudes, or the gas to liquid transition (red tide).  
Essentially, the understanding of the interaction between strings and 
membranes is important \cite{2f2,odi}.  
This is also intimately related to the string 
or membrane's condensation and the generation of Einstein gravity from 
the topological one.
\item From the estimation of the swimming efficiency, difference between 
the ciliate and the flagellate comes out.  In this respect the classification 
of the three kinds of swimming ways can be understood from the geometry of 
the shape space or its singularity structure. 
\end{enumerate}
\end{enumerate}

\end{document}